\documentclass[a4paper,11pt]{article}
\usepackage{pos}

\usepackage[utf8]{inputenc}

\usepackage{charter}
\usepackage{amssymb,amsmath,amsfonts}
\usepackage{mathtools}
\usepackage{mathrsfs}
\usepackage{bbm}
\usepackage{slashed}
\usepackage{nicefrac}

\usepackage{graphicx}

\usepackage{simplewick}

\usepackage{hyperref}
\usepackage{xparse}
\usepackage{xspace}

\usepackage{cancel}
\usepackage[normalem]{ulem}
\usepackage{xifthen}
\usepackage{dsfont}
\usepackage[titletoc]{appendix}
\usepackage{booktabs}
\usepackage{units}

\def\bs{\boldsymbol}

\long\def\comment#1{ }

\def\be{\begin{eqnarray*}}
\def\ee{\end{eqnarray*}}
\def\beq{\begin{eqnarray}}
\def\eeq{\end{eqnarray}}
\newcommand{\bea}{\beq \begin{aligned}}
\newcommand{\eea}{\end{aligned}\eeq}

\def\0{{\boldsymbol 0}}
\def\l{{\boldsymbol l}}

\def\rme{{\rm e}}

\def\dd{{\rm d}}

\def\tform{{t_\text{f}}}
\def\tdecoh{t_\text{d}}

\def\pT{p_{\scriptscriptstyle T}}

\def\out{\text{out}}
\def\in{\text{in}}

\DeclareDocumentCommand \TransAmp { o O{r} O{s} m m }
{\IfNoValueTF{#1}
	{T_{#2, #3} (#4, #5 )}
	{T^{(#1)}_{#2, #3}(#4, #5 )}
}
\DeclareDocumentCommand \ScalarPropMix { o m m m }
{\IfNoValueTF{#1}
	{\mathcal{G}_\text{\tiny F}(#3, #4 | #2 )}
	{\mathcal{G}^{(#1)}_\text{\tiny F}(#3, #4 |#2 )}
}
\DeclareDocumentCommand \ScalarPropMixTwo { o m m m m m }
{\IfNoValueTF{#1}
	{\mathcal{G}_\text{\tiny F}(#3, #5; #4, #6 |#2)}
	{\mathcal{G}^{(#1)}_\text{\tiny F}(#3, #5; #4,#6 | #2 )}
}
\DeclareDocumentCommand \ScalarPropConf { o m m m }
{\IfNoValueTF{#1}
	{\mathcal{G}_\text{\tiny F}(#2, #3 | #4 )}
	{\mathcal{G}^{(#1)}_\text{\tiny F}(#2, #3 | #4 )}
}
\DeclareDocumentCommand \ScalarPropRetMix { o m m m }
{\IfNoValueTF{#1}
	{\mathcal{G}(#3 , #4 | #2 )}
	{\mathcal{G}^{(#1)}(#3 , #4 | #2 )}
}
\DeclareDocumentCommand \ScalarPropRetMixTwo { o m m m m m }
{\IfNoValueTF{#1}
	{\mathcal{G}(#3, #5; #4, #6 | #2 )}
	{\mathcal{G}^{(#1)}(#3, #5; #4,#6 | #2 )}
}
\DeclareDocumentCommand \ScalarPropRetConf { o m m m }
{\IfNoValueTF{#1}
	{\mathcal{G}(#2, #3 | #4 )}
	{\mathcal{G}^{(#1)}(#2, #3 | #4 )}
}
\title{Cone-size dependent jet spectrum in heavy-ion collisions}
\author[a]{Yacine Mehtar-Tani}
\author[b]{Dani Pablos}
\author*[b]{Konrad Tywoniuk}

\affiliation[a]{RIKEN BNL Research Center and Physics Department, Brookhaven National Laboratory, Upton, NY 11973, USA}

\affiliation[b]{Department of Physics and Technology, University of Bergen, 5007 Bergen, Norway}

\emailAdd{mehtartani@bnl.gov}
\emailAdd{daniel.pablos@uib.no}
\emailAdd{konrad.tywoniuk@uib.no}
\abstract{
Jets in the vacuum correspond to multi-parton configurations that form via a branching process sensitive to the soft and collinear divergences of QCD. In heavy-ion collisions, energy loss processes that are stimulated via interactions with the medium, affect jet observables in a profound way. 
Jet fragmentation factorizes into a three-stage process, involving vacuum-like emissions above the medium scale, induced emissions enhanced by the medium length and, finally, long-distance vacuum-like fragmentation. 
This formalism leads to a novel, non-linear resummation of jet energy loss. In this talk we present new results on the combined effects of small-$R$ resummation and energy loss to compute the $R$-dependent jet spectrum in heavy-ion collisions. 
}

\FullConference{%
  HardProbes2020\\
  1-6 June 2020\\
  Austin, Texas}

\begin{document}
\maketitle

\section{Introduction}
Jets are collimated sprays of energetic particles that are copiously produced in high-energy collider experiments. They are proxies of perturbative partons, i.e. quarks and gluons, that are produced in high momentum-transfer scatterings and reveal their complex fragmentation patterns. In heavy-ion collisions, the space-time structure of jet fragmentation is particularly pertinent since it occurs in the background of a hot and dense QCD medium. Final-state interactions between jet constituents and the medium modifies the inclusive spectra as well as the intra-jet distributions \cite{dEnterria:2009xfs,Mehtar-Tani:2013pia,Blaizot:2015lma}.
% In these proceedings, we mainly focus on the class of observables belonging to the first category.

We present here ongoing work toward properly implementing a formalism of jet energy loss that a) accounts for the multi-prong structure of jets that interact with the plasma and their fluctuations and b) that correctly models how energy is transferred from the leading particle to large angles, where it ultimately thermalizes. In addition, the resummation of logarithms of the cone-size $R$, i.e. $\alpha_s \ln 1/R$, are consistently resummed using the standard DGLAP evolution equations \cite{Dasgupta:2016bnd}. These emissions correspond to large-angle vacuum-like emissions that are not captured by the jet cone.

%\section{Vacuum-like emissions in the medium and multi-parton quenching}
\section{From single parton quenching to multi-parton quenching}
For a single parton, produced according to the initial parton spectrum $\dd \hat \sigma/\dd \pT\sim \pT^{-n}$, we write the modified spectrum as
\beq
\label{eq:qf}
\frac{\dd\hat  \sigma_\text{med}}{\dd \pT} = \int_0^\infty \dd \epsilon \, P^{(0)}(\epsilon) \frac{\dd \hat \sigma_\text{vac}(\pT+\epsilon)}{\dd \pT'} \simeq Q^{(0)}(\nu) \frac{\dd \hat \sigma_\text{vac}}{\dd \pT} \,,
\eeq
where $\nu \equiv n/\pT$ and the quenching factor $Q^{(0)}(\nu) = \int_0^\infty \dd \epsilon \, \rme^{-\nu \epsilon} P^{(0)}(\epsilon)$.
This approximation relies crucially on the assumption of small energy losses, i.e. $\epsilon/\pT \ll 1$, and on the steeply falling nature of the spectrum, characterized by a large $n \gg 1$.
Since the quenching factor is the Laplace transform of the probability distribution, all sources of energy loss can be accounted for multiplicatively. 
%For the time begin, we only consider radiative processes.
For example, the energy loss induced via soft, in-medium radiation taking place off a highly energetic particle can approximately be treated as a Poissonian process. In this case, we obtain
\beq
\label{eq:qf-1}
Q_{\rm rad}^{(0)}(\nu) = \exp \left[- \int_0^\infty \dd \omega \, \frac{\dd I_>}{\dd \omega} \left(1- \rme^{-\nu \omega} \right) \right] \,,
\eeq
where $\dd I_>/\dd\omega = \int_{\bs k} \dd I / (\dd \omega \dd^2 {\bs k}) \Theta(|{\bs k}| -\omega R)$ is the spectrum of emission outside of the jet cone.
%For energy loss via elastic drag, we assume $P(\epsilon) = \delta(\epsilon - \hat e L)$ where $\hat e$ is the relevant transport coefficient. This leads to a quenching factor,
%\beq
%Q_{\rm el}^{(0)}(\nu) = \rme^{-\hat e L \nu} \,.
%\eeq

The generalization to multiple partons follows from the same approximations leading up to the factorized expression \eqref{eq:qf}. As an illustrative example, let us consider the exclusive $N$-parton production cross section, where $k_1 + \ldots + k_N = \pT$. Leaving details of the coherence and space-time  properties of the $\pT \to \{k_1,\ldots, k_N \}$ fragmentation process aside, we also assume that all $N$ partons are affected by medium energy loss. Since the most sensitive part of the process is the hard spectrum of the ``primordial'' parton carrying energy $\pT$ in the $n \gg 1$ limit, it is possible to show that
\beq
\label{eq:sigma-exclusive}
\frac{\dd^N \hat\sigma_{\rm med}}{\dd k_1 \ldots \dd k_N} = \Big[Q^{(0)}(\nu) \Big]^N \frac{\dd^N \hat \sigma_{\rm vac}}{\dd k_1 \ldots \dd k_N} \,.
\eeq
Hence, the main result of multi-parton energy loss is to apply an overall reweighting factor, which only depends on the total momentum, to the multi-parton system without producing further modifications of the details of the fragmentations process. These will appear as sub-leading contributions in the $\nu$ expansion.

A full treatment of a parton shower, including its coherence and space-time properties, can be realized as an implicit equation for a generating functional \cite{Dokshitzer:1991wu}. Considering the distribution of partons created in a fragmentation process, we separate three stages: {i)} vacuum-like radiation on time-scales shorter than the medium resolution time, {ii)} medium-induced processes affecting all resolved charges, and {iii)} further vacuum-like fragmentation after energy has been redistributed. The phase space available for emissions in stage {i)} is \cite{Mehtar-Tani:2017web}
\beq
({\rm PS})_{\rm ii)} = 2 \frac{\alpha_s C_R}{\pi} \ln\frac{R}{\theta_c} \left( \ln\frac{\pT}{\omega_c} +\frac{2}{3} \ln \frac{R}{\theta_c}\right) \,,
\eeq
with $\omega_s \sim \hat q L^2$ and $\theta_c \sim (\hat q L^3)^{-1/2}$ and $\hat q$ is the jet quenching parameter.
The impact of such emissions on the inclusive jet spectrum was first realized in \cite{Mehtar-Tani:2017web}. It was also interpreted in terms of a factorization of the in-medium branching process in \cite{Caucal:2018dla}. The second stage is encoded in the single-parton quenching factors, i.e. via the medium-induced spectrum in \eqref{eq:qf-1}.

In order to merge these three stages, we have to invoke two angular-ordered showers. Starting with the last stage, the late  ``out'' shower is a vacuum-like vetoed shower, i.e.
\beq
\label{eq:gf-out}
Z_\out(p,R) = u(p) + \int^R \dd \Pi \, (1-\Theta_\in) \left[Z_\out (zp,\theta) Z_\out ((1-z)p,\theta) - Z_\out(p,\theta) \right] \,,
\eeq
where $u(p)$ are test functions and the phase space $\dd \Pi \equiv \frac{\alpha_s}{\pi} \frac{\dd \theta}{\theta} p(z) \dd z$ involves the un-regularized Altarelli-Parisi splitting functions (we have suppressed the flavor indices) and we define the in-medium phase space as $\Theta_{\rm in} = \Theta( \tform < \tdecoh < L) $, with $\tform = 2/[z(1-z)p\theta^2]$ and $\tdecoh = (12/\hat q \theta^2)^{1/3}$. The final form of the generating functional is 
\beq
\label{eq:gf-full}
Z(p,R) = Z_\out(p,R')Q^{(0)}(\nu) + \int^R \dd \Pi \,\Theta_\in \left[Z (zp,\theta) Z ((1-z)p,\theta) - Z(p,\theta) \right] \,.
\eeq
It is worth discussing in detail all the terms in \eqref{eq:gf-full}. The right-most term on the right hand side accounts for multiple emissions inside the medium. The left-most term describes fragmentation outside of the medium, iterated by \eqref{eq:gf-out}. However, every resolved parton that escapes the medium brings an additional quenching factor $Q^{(0)}(\nu)$, similar to \eqref{eq:sigma-exclusive}. Furthermore, the ``out'' shower starts at a arbitrary large angle $R'\sim 1$, i.e. breaks angular ordering with respect to any previous emission. This is a consequence of the condition $\tdecoh < L$. Naturally, for intra-jet distributions $R'$ is fixed by the jet cone.

The normalization of the generating functional describes the suppression of the inclusive jet spectrum. We have that $Z(p,R)|_{\{u=1\}}= Q(p,R)$ where
\beq
Q(p,R) = Q^{(0)}(\nu) + \int^R \dd \Pi \,\Theta_\in \left[Q (zp,\theta) Q ((1-z)p,\theta) - Q(p,\theta) \right] \,.
\eeq
Here, the single-parton quenching factor $Q^{(0)}(\nu)$ acts as an initial condition for the complete resummed quenching factor of the jet.

\section{$R$-dependent jet spectrum in heavy-ion collisions}

We have computed the $R$-dependent spectrum in heavy-ion collisions. The initial hard spectrum was matched to PYTHIA8 simulations at $R=1$ with EPS09 LO nPDFs for LHC conditions and evolved using DGLAP evolution equations to the measured jet cone size. Medium parameters were fixed to $L=3.5$ fm and $\hat q = 1.5$ GeV$^2$/fm
%sampled from a realistic geometric distribution and propagated through a hydrodynamical background 
so as to describe existing data on jet suppression at $R=0.4$.  We have neglected elastic energy loss, and computed the medium-induced radiation spectrum in the BDMPS approximation assuming Gaussian transverse momentum broadening. 
%Further details can be found in \cite{}.
\begin{figure}
\centering
\includegraphics[width=0.5\textwidth]{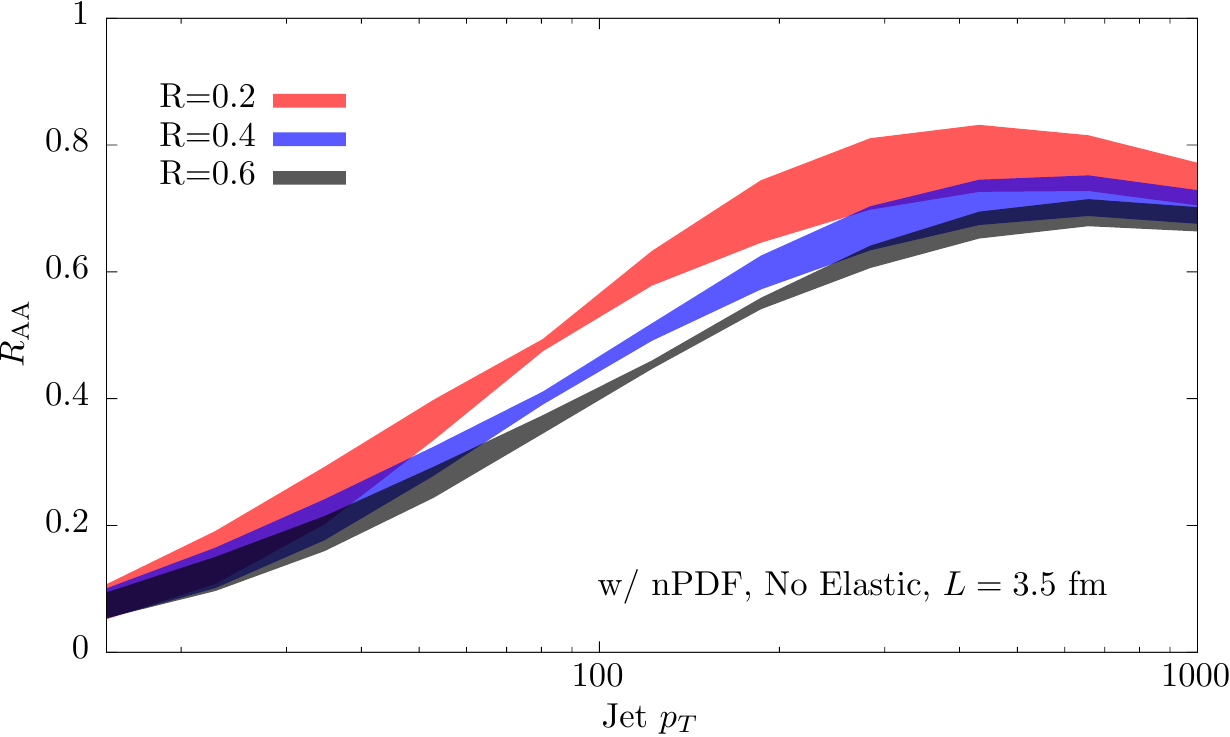}
\caption{Cone-size dependent jet nuclear modification factor $R_{AA}$ for heavy-ion collisions at the LHC. }
\label{fig:raa}
\end{figure}
The results presented at the conference are shown in Fig.~\ref{fig:raa}. 
%The parameters were chosen in order to reproduce the ATLAS data on jet suppression for $R=0.4$. 
The error bands correspond to varying the parameter $\theta_c$ by a factor of 2.	

We observe a weak $R$-dependence of the nuclear modification factor that can be interpreted as a consequence of color coherence: only splittings that occur at angles larger than $\theta_c$ will be resolved and affect the cone-size dependence of the spectrum. To summarize, jet energy loss is not only sensitive to the mechanisms of energy loss of single partons but also to how a jets as coherent quantum states interact with a thermal environment. The study of jet substructure observables in combination with the inclusive jet spectrum may allow to further constraint the theory in view of extracting medium transport properties. 

\paragraph{Acknowledgements:} KT and DP are supported by a Starting Grant from Trond Mohn Foundation (BFS2018REK01) and the University of Bergen. The work of Y. M.-T. was supported by the U.S. Department of Energy, Office of Science, Office of Nuclear Physics, under contract No. DE- SC0012704, by Laboratory Directed Research and Development (LDRD) funds from Brookhaven Science Associates and by the RHIC Physics Fellow Program of the RIKEN BNL Research Center.

\bibliographystyle{JHEP}
\bibliography{hp2020}

\end{document}